\documentclass[conference]{IEEEtran}
\IEEEoverridecommandlockouts
\usepackage{cite}
\usepackage{amsmath,amssymb,amsfonts}
\usepackage{algorithmic}
\usepackage{graphicx}
\usepackage{textcomp}

\usepackage{amssymb}% http://ctan.org/pkg/amssymb
\usepackage{pifont}% http://ctan.org/pkg/pifont

\usepackage{xcolor}
\def\BibTeX{{\rm B\kern-.05em{\sc i\kern-.025em b}\kern-.08em
    T\kern-.1667em\lower.7ex\hbox{E}\kern-.125emX}}
\begin{document}

\title{Review of Peer-to-Peer Botnets and Detection Mechanisms}
\author{\IEEEauthorblockN{1\textsuperscript{st} Khoh Choon Hwa}
\IEEEauthorblockA{\textit{National Advanced IPv6 Center} \\
\textit{Universiti Sains Malaysia}\\
Penang, Malaysia}
\and
\IEEEauthorblockN{2\textsuperscript{nd} Selvakumar Manickam}
\IEEEauthorblockA{\textit{National Advanced IPv6 Center} \\
\textit{Universiti Sains Malaysia}\\
Penang, Malaysia\\
selva@usm.my}
\and
\IEEEauthorblockN{3\textsuperscript{rd} Mahmood A. Al-Shareeda}
\IEEEauthorblockA{\textit{National Advanced IPv6 Center} \\
\textit{Universiti Sains Malaysia}\\
Penang, Malaysia\\
alshareeda022@gmail.com}

% \and
% \IEEEauthorblockN{4\textsuperscript{th} Given Name Surname}
% \IEEEauthorblockA{\textit{dept. name of organization (of Aff.)} \\
% \textit{name of organization (of Aff.)}\\
% City, Country \\
% email address or ORCID}
% \and
% \IEEEauthorblockN{5\textsuperscript{th} Given Name Surname}
% \IEEEauthorblockA{\textit{dept. name of organization (of Aff.)} \\
% \textit{name of organization (of Aff.)}\\
% City, Country \\
% email address or ORCID}
% \and
% \IEEEauthorblockN{6\textsuperscript{th} Given Name Surname}
% \IEEEauthorblockA{\textit{dept. name of organization (of Aff.)} \\
% \textit{name of organization (of Aff.)}\\
% City, Country \\
% email address or ORCID}
}

\maketitle

\begin{abstract}
Cybercrimes are becoming a bigger menace to both people and corporations. It poses a serious challenge to the modern digital world. According to a press release from 2019 Cisco and Cybersecurity Ventures, Cisco stopped seven trillion threats in 2018, or 20 billion threats every day, on behalf of its clients. According to Cybersecurity Ventures, the global cost of cybercrime will reach \$6 trillion annually by 2021, which is significantly more than the annual damage caused by all natural disasters and more profitable than the global trade in all major illegal narcotics put together. Malware software, including viruses, worms, spyware, keyloggers, Trojan horses, and botnets, is therefore frequently used in cybercrime. Especially for businesses that heavily rely on digital information technologies, it has become crucial for businesses across all industries and governments to identify and stop cyberattacks before they cause significant damage. The most common malware employed by attackers to carry out cybercrimes is the botnet, which is available in a variety of forms and for a variety of purposes when attacking computer assets. Botnets are the most common and have a significant negative impact on any civilization among malware programmes. It is difficult to identify and stop malicious behaviours from occurring without a thorough grasp of the relevant cybercrime operations, such as malicious botnet activities. Fortunately, a wide range of strategies have been created and advocated to tackle the botnet issue. However, the issue continues to exist and worsen, seriously harming both enterprises and people who conduct their business online. The detection of P2P (Peer to Peer) botnet, which has emerged as one of the primary hazards in network cyberspace for acting as the infrastructure for several cyber-crimes, has proven more difficult than regular botnets using a few existing approaches. As a result, this study will explore various P2P botnet detection algorithms by outlining their essential characteristics, advantages and disadvantages, obstacles, and future research.

\end{abstract}

\begin{IEEEkeywords}
 P2P botnet, detection mechanism.
\end{IEEEkeywords}

\section{Introduction}

The majority of Internet users utilise the Internet to conduct many of their usual daily activities, including routine work, social interactions, and personal leisure. Nearly all Internet users claim that their everyday activities and routines would be altered if they lost access to the Internet. Our lives have altered in every way because to the internet and communication technologies. Internet has a lot to offer us in terms of valuable services and applications. However, security risks, often known as cybercrimes that affect people and companies, are one of the biggest downsides of the Internet \cite{baruah2019botnet}. Malicious software, sometimes known as malware, has thus risen to prominence in today's high-tech society. Malwares are made up of a wide range of programmes, including Trojans, worms, spyware, keyloggers, and botnets. The most pervasive and significant threat to computing assets among the various types of malicious software is posed by botnets. The increasing dependence on the internet over the past few decades has made it difficult to manage the integrity, confidentiality, and security of user data and computing resources \cite{1,2,3,4,5,6,7,8,9,10,11,12,13,14,15,16,17,18,19,20,21,22,23,24,25}. This is due to the fact that the majority of cyber-security-related problems are caused by malicious software that is operating undetected on user computers and may jeopardise the security of the user's data, especially in critical infrastructure, government, business, and academic settings\cite{khehra2018botnet}. 

A botnet is a group of compromised computers that the attackers covertly control and utilise for a variety of harmful purposes. A bot master, who controls a botnet's command and control for remote process execution, is the attacker. The infiltrated computers or other devices in a botnet are known as zombies or bots \cite{rawat2018survey}. Using a set-up C\&C channel, Botmaster has the capacity to remotely control the behaviour of bot malware, making bot operations more adaptable and customising instructions to suit its requirements. The bot master uses the botnet he controls to carry out a variety of malicious activities, including Distributed Denial of Service (DDoS), sending spam emails with viruses attached, phishing, spreading malware, cracking passwords, stealing identities, committing internet fraud, key logging, and extorting online businesses, among others. The bot master needs powerful computational power in order to carry out the nefarious acts. Thus, a bot master makes various social engineering and exploiting attempts to attack weak computers or devices (systems with fewer protection mechanisms) on a network. Once the computers or devices are infected, the legitimate owner or user will not even be aware of it, and without his or her consent, the computers or devices will become a member of their botnets, where the personal data and credentials that are stored there will be stolen or they will engage in other malicious activities as directed by the bot master\cite{baruah2019botnet}. 

This paper focuses solely on discussing numerous P2P botnet topics, including P2P botnet design, C\&C communication, P2P detection methods, and observations and difficulties. The following are some of its noteworthy contributions:

\begin{itemize}
    \item P2P botnets' life cycle of development is discussed. 
    \item It presents the taxonomy and thorough analysis of the various detection frameworks and models. 
    \item The research observation and difficulties in using these P2P botnet detection methods are discussed.
\end{itemize}

The reminder of this paper is organized as follows. Section \ref{resec} covers the background and related surveys. The various P2P detection techniques are discuss in Section \ref{secpot}. In Section \ref{secs}, it will presenting the identified research observation and challenges. Finally the Section \ref{sec:con} presents the summary and the directions for future research work.

\section{Background and Related Survey}\label{resec}

A group of infected computers that are managed by the bot master via command and control (C\&C) channels is referred to as a botnet. A botnet usually consists of three different kinds of software. As follows:

\begin{itemize}
    \item Server program: To control infected computers or bots, these programmes are placed on the command and control (C\&C) server \cite{baruah2019botnet}.
    
    \item Client program: To control infected computers or bots, these programmes are placed on the command and control (C\&C) server \cite{baruah2019botnet}. 
    
    \item Malicious program: These applications or programmes fall under the category of malware since they are used to infect or compromise susceptible machines online. Malware like Gnuman and VBS that is managed by a bot master has been found to compromise computers. Trojan.Peacomm, Gnutella, SdDrop, and others \cite{baruah2019botnet}. 

\end{itemize}

The botnet's communication system is another crucial element. Bots are always in communication with the C\&C server, receiving instructions to engage in nefarious activity. The bots then continuously listen for commands, carry out the tasks as directed, and backup the acquired data to the C\&C server. They are divided into three groups based on how the botnets communicate: IRC botnets, HTTP botnets, and P2P botnets.

\subsection{IRC Botnet}

The earliest botnet in existence is the IRC botnet. They are not classified as malicious bots in the design of these bots. In actuality, the majority of bots are useful and necessary for the operation of the Internet. Internet Relay Chat (IRC) protocol chatroom functioning was facilitated by the first internet bots. 

Search engines utilise web crawler internet bots for indexing and updating. In addition to indexing websites, bots are also used to gather current news and price data. Chat bots are an additional useful area. Siri is the most well-known chatbot, and they are utilised as virtual assistants and in social services. Additionally, there is a moral and legal grey area surrounding the use of commercial internet bots for automated trading, auction bidding, and monitoring of product positioning and reviews \cite{wainwright2019analysis}. The IRC protocol was first created for extensive data distribution and conversation among end users, as shown in the examples above. The IRC protocol's inherent flexibility and scalability have been used by criminals to carry out a variety of nefarious activities. IRC botnets are regarded as having a centralised C\&C architecture. In other words, a single C\&C server is in charge of all the bots. (View Figure \ref{dderes33d}) As a result, it is susceptible to single point of failure due to the centralised architecture. The entire botnet will disintegrate as soon as the server is identified and taken offline \cite{baruah2019botnet}.

\subsection{HTTP botnet}
It is comparable to the IRC botnet in this HTTP (Hypertext Transfer Protocol) botnet. The bot masters set up HTTP servers so that they may communicate with the infected computers using the HTTP protocol. Due to the popularity of the HTTP protocol on the Internet, its use by many apps, and its ease of entrance toolkits, bot masters aim to blend their malicious bot traffic with legitimate HTTP traffic in order to avoid detection. The HTTP botnet also falls under the category of centralised C\&C architecture, and it can be developed into a hierarchical design with specific subgroups of bots structured for load balancing and for specific content distribution, such as spam \cite{wainwright2019analysis}. The centralised C\&C structure is reliable and simple to put into practise. Bot master has the capacity to tell all bots to react, respond, and attack simultaneously over communication. Due to the fact that the C\&C server contained information about the whole botnet, a single point of collapse resulted \cite{zhuang2018enhanced}.

\subsection{P2P botnet}
\begin{figure}[h]
	\includegraphics[width=.45\textwidth]{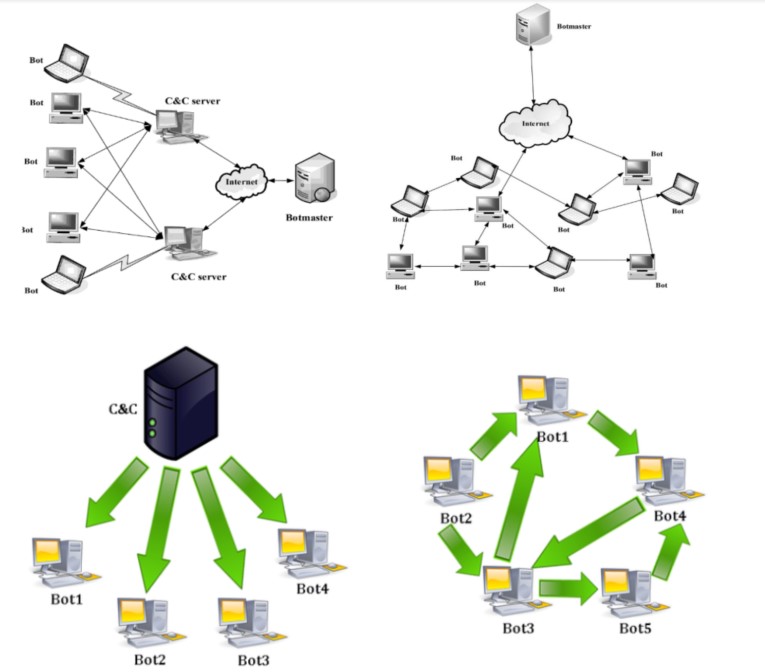}
	\caption{A typical Centralized Botnet VS Peer to Peer Botnet Architecture.} \label{dderes33d}
\end{figure}

If the C\&C server is discovered, the entire botnet could be destroyed, as is understood with the IRC and HTTP botnets. As centralised botnet mitigation became commonplace, botnets advanced to a decentralised C\&C architecture (See Figure \ref{dderes33d}). Instead of being connected to a central C\&C server, the bots with a decentralised architecture are connected to the neighbouring bots in the botnet. Because of this, the decentralised botnet is frequently referred to as a peer-to-peer (P2P) botnet. This structure is similar in principle to the peer-to-peer network model. The P2P architecture eliminates the vulnerability of a single point of disruption because every bot in the P2P botnet can act as either the client or server, but it increases the cost of reaction time \cite{baruah2019botnet}. The most recent botnet architecture makes use of social media sites like Twitter and Facebook, therefore the term "botnet" now refers to an automated social media account rather than a hijacked computer or other device. The bot master can then manually create numerous accounts or, for increased efficiency, register to the social media site using an application. Since social media sites don't charge for account registration, social media botnets often have cheap entrance and upkeep costs. The owner of a social media bot will use this opportunity to post malicious links, gather information on prominent targets, and disseminate information with the aim of influencing politics and business.

\subsection{Botnet Development Lifecycle}
\begin{figure}[h]
	\includegraphics[width=.45\textwidth]{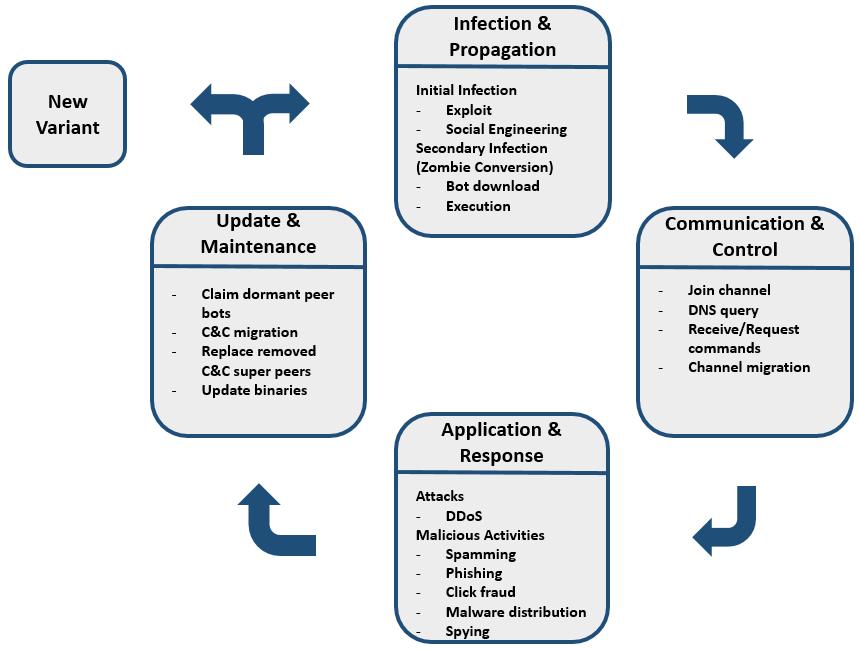}
	\caption{Botnet Development Lifecycle.} \label{dderes3botn3d}
\end{figure}

A product development model is employed to provide a taxonomy and lifecycle for botnets. The product development lifecycle of a botnet has numerous stages, including infection and propagation, communication and control, application and reaction, and update and maintenance. Below is a detailed explanation of each phase (See Figure \ref{dderes3botn3d}) \cite{wainwright2019analysis,rawat2018survey}.

\begin{itemize}
    \item Infection \& propagation. The motivation of the bot master is a major factor in whether or not they will construct a botnet. Money, entertainment, ego, and other factors are among the motivations that are described above. The bot master's goal at this point is to increase the number of nodes in the botnet. It can boost the dissemination by taking advantage of flaws in any of the many malware attack vectors, including websites and browsers, email, and social media platforms \cite{wainwright2019analysis}. The initial infection of the vulnerable nodes progresses to secondary infection (Zombie conversion), which spreads to the neighbouring nodes.
    
     \item Communication \& control. At this level, communication between internal and external botnets as well as potential vulnerability nodes is involved. It's crucial to distinguish botnets from other malware types by their communication capability. For connecting freshly infected nodes to the botnet channel, internal communication relies on the C\&C architecture, whilst external communication uses the Domain Name Service (DNS) to request the resolution of IP addresses. External communication is also in charge of the registration procedure that allows a vulnerable device to join the botnet. defences to avoid detection during the communication phase. so that the weak nodes can conceal the identity of the infected node.
      \item Application \& Response. By leasing the botnet's services or selling the code, the botnet can become profitable once the potential motivation level in terms of economic value is met. At this stage, botnet attacks are launched, including DDoS, spam, phishing, click fraud, virus distribution, espionage, and others.

       \item Update \& maintenance. Updates and maintenance, such as claiming dormant peer bots, are needed for the infected nodes that joined the active botnet. In order to strengthen the prevention strategy, the executable C\&C server may need to migrate, replace the C\&C super peers that were removed after being discovered, and update the binaries on all afflicted nodes. If a botmaster detects a botnet as a whole at this point, the development cycle must be restarted. If not, new botnet variants will continue to strike.
\end{itemize}

\subsubsection{Command and Control Mechanisms}
The bots are instructed by distant systems through the command and control (C\&C) channels. Based on the C\&C protocols, multiple C\&C architectures can be categorised, including IRC, HTTP, P2P, Bluetooth, email, social networks, DNS, and other unique protocols. The botnet is based on the demands that either P2P or non-P2P protocols be used for C\&C communication. The pull and push mechanisms in botnets are the 2 main C\&C activities. When using a pull system, bot masters post instructions at predetermined sites, where peer bots can subscribe and actively accept them. The bot subsequently follows the instructions and also notifies its list of neighbouring peers \cite{rawat2018survey}. To execute the commands, C\&C servers push or advance them to their neighbouring peers via the push technique. The nearby peer bots then wait passively for the instructions and also transmit the instructions they get to further nearby peer bots.

\section{P2P Botnet Detection Techniques}\label{secpot}

To categorise the attacks, the common botnet detection methods can be divided into three categories. (1) Botnet traffic, (2) Command and Control (C\&C) servers, and (3) PCs with bot infections. Nowadays, a lot of proposed detection schemas are based on one or more of the aforementioned criteria. This study groups the numerous ideas into the following subcategories based on these three characteristics: traffic-based detection, behavior-based detection, DNS-based detection, graph-based detection, data mining-based detection, soft computing-based detection, and general framework. More information on each detection approach will be provided in the next section.

\subsection{Traffic-based Detection}
The P2P bots interact with a large number of other peer bots to carry out push/pull mechanisms, provide the bots instructions to gather data, or deliver updates from the bot master to infected PCs. Consequently, it continually produces anomalous network traffic \cite{rawat2018survey}. For the objective of observing network patterns and monitoring network traffic in order to identify the presence of botnets, a number of traffic-based detection approaches have been presented. Multi-phased flow model is one of the well-known P2P traffic-based detection techniques. When acting as the sole peer in a P2P botnet, a bot must establish connections with as many neighbour peers as possible in order to build bot networks. As a result, the outcome illustrates the features of processing a large amount of network data in order to perform peer discovery and information exchange with several linked neighbour peers. In other words, it suggests that the network traffic flow resembles the network pattern and happens frequently at erratic intervals. For instance, a massive amount of TCP and UDP traffic created by P2P botnets is flown, analysed for correlations, and compressed into clusters that represent each flow phase in the assault traffic. As a result, by developing the modelling in a Markov chain framework, the traffic flows are grouped into models based on the states \cite{kwon2016psybog}.

The clustering of TCP or UDP communication formats creates the grouping and tracks packets to identify whether they are flooding attacks from botnets or regular network traffic transmission. Additionally, this method employs an algorithm to create a matrix of flow modelling and detecting engine based on network traffic. This method, however, can only identify P2P C\&C traffic that resembles user traffic while the model is being trained. In order to avoid discovery, malicious software or botnets may use network traffic patterns that are similar to those of genuine P2P networks.

The researcher also suggested a different method based on the flow dependence of C\&C network data to identify P2P botnets. By assuming that typical network traffic has complex short-term network flow dependence, this method distinguishes P2P networks from conventional P2P application network traffic \cite{rawat2018survey}. This method likewise relies on looking into the well-known network flow relationships. This approach may have trouble identifying network flow dependencies if the network flows haven't happened frequently in the past. Additionally, a lot of trace network sample patterns that represent synthetic P2P botnet traffic must be collected for this technique to function fairly. As a result, this technique is not excellent for scaling because it requires a lot of labour to rebuild the modelling\cite{rawat2018survey}.

\subsection{Behavior-based detection}
The behavioural characteristics of the botnet have been thoroughly analysed. Bots typically have a wide range of common features, including structured behaviour, maintaining consequence connections to communicate with and respond to neighbouring bots, as well as receiving commands from the bot master via a C\&C server. This method looks at the P2P botnets' behaviour and network characteristics, which are thought to be very closely related to their fundamental architecture and mode of operation. This technique focuses on the network behaviour that is done by the botnet after getting the order from the C\&C server will behave unlikely to be human behaviour rather than analysing the network traffic flow as per network-based detection.

One of the suggested behavior-based detection methods for identifying P2P botnets uses a list of behaviour metrics that ultimately derive from characteristics of standard network traffic, such as statistics on traffic pattern, topological properties, and protocol sequence, to identify hosts with similar connection patterns. The monitoring network must have several infected bots for this strategy to be effective. Additionally, threshold attacks launched by bot masters can avoid the thresholding metrics filtering in the list of behaviour metric attributes.

\begin{table}[h]
	\caption{Summary of detection techniques}
	\centering
	\renewcommand{\arraystretch}{1.3}
 	\begin{tabular}{p{35pt}p{40pt}p{50pt}p{30pt}p{40pt}}
		\hline
Detection Techniques & Detection Type&Network and structure bots & Real time Bots & Accuracy\\
           \hline
       Traffic-based detection    & Known & P2P & Yes& Detect P2P bots only in a monitored network\\
           Traffic-based detection &	Known	&P2P&	Yes&	Detect P2P bots only in a monitored network\\
Behavior-based detection &	Known &	P2P, structured	& Yes	& Detect P2P bots only in a monitored network\\
            DNS-based detection	Both &	P2P&	Yes&	Detect P2P bots only in a monitored network\\
           Graph-based detection &	Both&	P2P, Structured&	Yes&	Detect P2P bots only in a monitored network\\
           Data Mining-based detection &	Known and detected any new detection type &	P2P& Yes&	Detect P2P bots which feature had been identified before\\
           Machine learning-based detection	& Both &	P2P	& Yes	&High\\
           Generic Frameworks & Known	&P2p	&Limited	&Low\\
            \hline
	\end{tabular}
	\label{tab1summ11}
\end{table}

\begin{table*}[t]
	\caption{Strengths and limitation of P2P botnet detection techniques}
	\centering
	\renewcommand{\arraystretch}{1.3}
 	\begin{tabular}{p{45pt}p{220pt}p{190pt}}
		\hline
Detection Techniques & Strengths &Limitation and Challenges \\
            \hline
           Traffic-based detection & 
           \begin{itemize}
               \item Behavior \& Traffic Analysis, Multi-phased ﬂows model
\item C\&C Traffic detection
\item Flow dependencies, Independent of malicious Traffic
\item Structure \& protocol independent, Pattern based features
\item Real-time \& large scale	

           \end{itemize} & 

\begin{itemize}
\item Not detected by using a legitimate P2P network
\item Higher false positive
\item Not detected by blended peer bots and randomization
\item Not detected on traffic tunneling through Tor network
\item Detect P2P bots only in a monitored network 

\end{itemize}
\\

Behavior-based detection &
\begin{itemize}
  
  \item Exploit Traﬃc pattern
  \item Bots group behavior
  \item Host-network cooperation, Independent of topology \& protocol
  \item Resilient to encryption \& obfuscation
  \item Temporal resource sharing mode
  \item Monitoring resource sharing behavior
 
\end{itemize}

& \begin{itemize}
   	\item Multiple bots dependency, Vulnerable to threshold attack
\item Evasion by bots: using benign domains
\item Used only for parasite P2P botnets
\item Source should be popular and short life
  
\end{itemize}\\

DNS-based detection	& \begin{itemize}
    \item Group Activity Detector, Online unsupervised known, Unknown
    \item Scalable, Real-time
\end{itemize}
 & \begin{itemize}
     \item Requires multiple bots
 \end{itemize}\\

Graph-based detection &	
\begin{itemize}
 \item  Reachability \& centrality properties
 \item  C\&C channels detection, Monitoring bot activities
 \item  C\&C patterns in overlay topology
 \item  Large-scale, Clustering techniques
\end{itemize}
&
\begin{itemize}
\item  Vulnerable to random delay
\item  P2P protocols dependency, False negatives
\item Bootstrap information required
\end{itemize}\\

Data Mining-based detection &
\begin{itemize}

\item Mining Concept-Drifting Data Stream
\item Packet features are extracted and aggregated into Flow characteristics
\item Analysis of traffic features – Fingerprint botnet C\&C channels
\item Created application proﬁle from known P2P applications
\item Based on high-level statistical traffic features	

\item 
\end{itemize}
& 
\begin{itemize}
  
\item  Requires monitoring traffic at each host
\item Sampling may miss useful communications patterns
\item Evasion by random message padding
\item Dependency on the dialog-like pattern
\item Deals with the signaling flows as a whole
\item Evasion by randomization of inter-packet delays
  
\end{itemize}
\\

Soft computing-based detection &

\begin{itemize}
    
\item Traffic behavior, Detection in C\&C phase
\item Detection rate 98\%
\item Anomalous Network traffic
\item Real-time detection in C\&C phase \& attack phase
 
\end{itemize}
& 

\begin{itemize}
 \item  Dependency on features selection
 \item  High computational requirement
 \item Sampling can skip botnet flows
 \item Vulnerable to obfuscation
\end{itemize}\\

Generic Frameworks &

\begin{itemize}

 \item Anomaly-based-behavior, traffic-based analysis
 \item  Independent to protocol and C\&C structure, Realtime
 \item  Network traffic, Bot behavior, Detect Bots
 \item  No prior information required
 \item  Remote control process- analysis
 \item Active-informed probing, Fast, Scalable, Real time

\end{itemize}
&
\begin{itemize}
   
\item Detect only active bot(s)
\item Targets enterprise network only
\item Threshold attack
\item Content analysis required
\item False positives advanced encryption
\item Delayed port binding
 
\end{itemize}\\

            \hline
	\end{tabular}
	\label{tab1sutopttmm11}
\end{table*}

\subsection{DNS-based Detection}
The fundamental characteristic of the bots is a cluster of activity, and they frequently use the domain name system (DNS) to rearrange C\&C servers, update their bots' code, and conduct attacks. On rare occasions, the same DNS's bot traffic—which is distinct from that of actual users—dominates the DNS traffic \cite{rawat2018survey}. In terms of security, DNS traffic can be a rich information source. The majority of these DNS-based detection methods enable the identification of infected computers only based on the network traffic created by botnets. Untrustworthy computers can then be examined throughout the botnet lifespan. These methods can identify botnets that have not yet launched an assault and may still be in the formation stage. Unfortunately, as DNS traffic volume grows tremendously, security network administrators and analysts must contend with the difficulty of gathering, retrieving, and analysing DNS traffic in order to respond to contemporary Internet threats. In other words, the majority of DNS research is detrimental to this field. The DNS-based anomaly detection methods are described and assessed in this study \cite{ostap2017concept}.

This study suggested Botnet Group Activity Detector, an online unsupervised botnet detection method (BoTGAD)\cite{chang2015measuring}. BotGAD is only concerned with DNS flow analysis, with which it attempts to identify coordinated bot activity. A number of sensors have been put in various parts of the network to identify such activity, and bots use the DNS protocol to resolve the address of potential threats' domain names. All of the devices connected to the studied network will be recognised because the BotGAD is designed to only detect synchronous action. In contrast to legitimate network traffic, the bot master typically transmits commands to all of his bots. For those bots that get the commands, they attempt to send back the desired results of these commands as soon as feasible. As a result, the results are transferred quickly. Additionally, all running bots will deliver results in a synchronous way, which will increase the resemblance between each network session set up for bots to communicate with each other's masters. For example Elastic Zombie \cite{alonso2018data} and Blackshaded \cite{chang2015measuring} botnets sends control packet to C\&C conﬁrming activity at speciﬁed time, response time are 30 seconds. and 45 seconds respectively. As a result, synchronous activity can be seen during a variety of botnet operations. BotGAD, in contrast, is a good method but it also has a number of shortcomings. The main drawback of BotGAD is that it must divide a complete activity into time-windows with fixed lengths, which denote predefined and equal portions of the entire monitoring time. This is in addition to the fact that it only examines DNS protocol. In those windows, all source IPs of the examined destination IP are consolidated. The effectiveness of the group activity detection is significantly impacted by how the time window's length is determined.

\subsection{Graph-based detection}

Numerous research use various graph-based features to find network anomalies. The spatial relationships are mostly used by the graph-based features to understand the botnet network activity. The characteristic patterns of the botnets can be discovered through the graphical analysis of the botnet communication interaction. BotGrep is a well-known example of a graph-based detection. The examination of network flows gathered over numerous big networks, such as ISP networks, is how the BotGrep finds P2P botnets. In the BotGrep concept, a P2P network is first recognised as a group of hosts in the overall view of Internet traffic. The structure of Kademlia-based P2P botnets, as seen from a graph theoretical perspective and as implemented as C\&C communications, heavily influences the algorithm. To bootstrap the detecting algorithm, BotGrep requires additional data. It may be difficult to maintain an overall perspective on Internet communications when bootstrapping the detection algorithm.

\subsection{Data Mining-based detection}
Based on data mining techniques, various research projects have been proposed to identify malicious activities carried out by botnets. These methods are capable of categorising actual unseen threat samples, recognising the threat families of malicious samples, and deducing the feature. In essence, these strategies involve the two steps of feature extraction and categorization. Binary strings, API calls, and programme actions are just a few of the features that are statistically extracted as the key features in the feature extraction process. These features capture the characteristics of the file samples dynamically. Once the feature extraction process's data collecting is complete, the next step, known as classification, can be taken.

In this stage, the file samples are classified into appropriate groups or classes based on the examination of feature categories produced by the feature extraction process, using intelligent approaches like classification or clustering \cite{rawat2018survey}. As a result, the primary differences between these data mining-based detection strategies relate to the feature category and the data mining techniques used. Effective threat detection, whether from known or unidentified botnets, heavily depends on the quality of the trained model. The model must first be built and trained by the system before it can be utilised to make inferences. Without a high-quality training sample, the system cannot be trained with good software applications or malware. In order to extract the features and comprehend the underlying qualities, each sample is parsed. The training set then included the conversion of all the extracted features into digital vectors with labels for each feature, or the so-called trained model.The classification algorithm is then given the trained model to create the classifiers. Depending on the learned model that is utilised in the system, the classification algorithm can either identify a sequence of unknown input file samples as dangerous or legitimate applications.

\subsection{Machine learning-based detection}
Based on network traffic, network activity, and a number of additional features for study, there are numerous approaches to identify P2P botnets. This section suggested a method for checking the flexibility, inventiveness, and early detection of P2P botnets using machine learning. One of the suggested methods for achieving real-time detection is to have the system analyse the properties of network traffic flows in brief time frames. Utilizing a botnet architecture with supervised machine learning algorithms is another method. This method uses a framework that extracts conversation-based features via learning from random forests \cite{rawat2018survey}.

\subsection{Generic Frameworks}
Additionally, other general frameworks for detecting botnets have been put out and are based on traffic correlation analysis and behaviour tracking. One common framework for locating the botnet is BotMiner. This method is applicable on a small scale and does not scale well since it relies on network packets and flow analysis, which takes a lot of fine-grained data to study the network \cite{rawat2018survey}.

\section{Observations and Challenges}\label{secs}

Botnet detection methods are compared based on their performance in identifying known and unknown bots, protocol and structure bots, encrypted C\&C channel bots, real-time bots, and accuracy. This comparison are summarize into Table \ref{tab1summ11} and Table \ref{tab1sutopttmm11} below:

\section{Conclusion}\label{sec:con}

This research provided a thorough analysis of several facets of P2P botnets. There is no one detection method that can consistently identify evolving botnets because each detection method has its own strengths, weaknesses, and scope. Furthermore, the majority of detection techniques rely on offline analysis, grouping, and classification, and as a result, do not take into account the needs of real-time detection. As a result, the need to build a real-time detection technique for clustering and categorising botnet traffic as well as on-the-fly mining of the botnet traffic is now very important.

	\bibliographystyle{unsrt}
	\bibliography{ref}
\end{document}